\documentclass{JHEP3}
\pdfoutput=1
\usepackage{amsmath,amssymb,epsfig}

\newcommand{\be}{\begin{equation}}
\newcommand{\ee}{\end{equation}}
\newcommand{\beq}{\begin{equation}}
\newcommand{\eeq}{\end{equation}}
\newcommand{\ba}{\begin{eqnarray}}
\newcommand{\ea}{\end{eqnarray}}
\newcommand{\bea}{\begin{eqnarray}}
\newcommand{\eea}{\end{eqnarray}}

\def\cH {{\mathcal{H}}}
\def\cJ {{\mathcal{J}}}
\def\cK {{\mathcal{K}}}

\def\pa    {{\partial}}

\def\L {\Lambda}
\def\m {\mu}

\def\o {\omega}
\def\p {\phi}
\def\P {\Phi}

\def\vt#1#2#3 {{\vartheta[{#1 \atop  #2}](#3\vert \tau)}}

\newcommand{\poi}[2]{\{#1,#2\}}

\title{On the addition of torsion to chiral gravity}

\author{Ricardo Couso Santamar\'\i a\\
\sl Department of Particle Physics and IGFAE, University of Santiago de Compostela, E-15782 Santiago de Compostela, Spain\\\vskip-4mm
\email{ricardo.couso@usc.es}}

\author{Jos\'e D. Edelstein\\
\sl Department of Particle Physics and IGFAE, University of Santiago de Compostela, E-15782 Santiago de Compostela, Spain\\\vskip-3mm
\sl Centro de Estudios Cient\'{\i}ficos, CECS\\
Casilla 1469, Valdivia, Chile
\\\vskip-4mm
\email{jose.edelstein@usc.es}}

\author{Alan N. Garbarz\\
\sl Physics Department, University of Buenos Aires, FCEN-UBA\\
Ciudad Universitaria, Pabell\'{o}n 1, 1428, Buenos Aires, Argentina\\\vskip-4mm
\email{alan@df.uba.ar}}

\author{Gast\'on E. Giribet\\
\sl Institute of Physics of Buenos Aires, IFIBA-CONICET\\
Ciudad Universitaria, Pabell\'{o}n 1, 1428, Buenos Aires, Argentina\\\vskip-4mm
\email{gaston@df.uba.ar}}

\bigskip
\bigskip
\abstract{Three-dimensional gravity in Anti-de Sitter space is considered, including torsion. The derivation of the central charges of the algebra that generates the asymptotic isometry group of the theory is reviewed, and a special point of the theory, at which one of the central charges vanishes, is compared with the chiral point of topologically massive gravity. This special point corresponds to a singular point in Chern-Simons theory, where one of the two coupling constants of the $SL(2,\mathbb{R})$ actions vanishes. A prescription to approach this point in the space of parameters is discussed, and the canonical structure of the theory is analyzed.}

\preprint{}

\begin{document}

\section{Introduction}

The model of chiral gravity proposed by Li, Song and Strominger in \cite{Li2008,Strominger2008} represents a very interesting idea to construct a consistent theory of quantum gravity in three-dimensions. The feasibility of formulating such a model was extensively discussed in the last three years \cite{Carlip2008,Carlip2009,Li2008a,Giribet2008b,Carlip2008c,Carlip2009c,Henneaux2009,Grumiller2008,Grumiller2009,Ertl2009,Maloney2010} and is still matter of technical analysis \cite{Henneaux2010,Gaberdiel2010,Cunliff2010,Compere2010}. Here, chiral gravity is discussed and compared with a special (singular) limit of Chern-Simons gravity.

We consider the most general Chern-Simons gravitational theory in tree-dimensional Anti-de Sitter space (AdS$_{3}$), including torsion \cite{Mardones1991,Mielke1991}. We briefly discuss how the central charges of the (conjecturally existing \cite{Maldacena1998,Witten1998,Gubser1998}) dual conformal field theory (CFT)\ can be calculated. This can be done by implementing AdS$_{3}$ asymptotic boundary conditions, both for the dreibein and for the spin connection, which amounts to perform the Hamiltonian reduction of the boundary action, straightforwardly adapting what is known for the case of three-dimensional Einstein gravity in AdS$_{3}$ \cite{Coussaert1995,Henneaux2000}. The result consistently agrees with the central charges previously obtained in the literature by different methods \cite{Blagojevic2003a,Blagojevic2003,Blagojevic2003b,Klemm2008,Cacciatori2006}.

It is observed that the theory exhibits a special point in the space of parameters, at which it becomes chiral by construction as one of the two coupling constants of the $SL(2,\mathbb{R})$ Chern-Simons actions vanishes. This is a singular point of the Chern-Simons theory, and this singularity was recently mentioned in \cite{Witten2010} in the context of the analytically extended theory, where a relation between this singular point and the chiral point of \cite{Li2008} was already pointed out. The fact that at this point one of the two $SL(2,\mathbb{R})$ Chern-Simons actions (say the left-handed one) decouples implies that the degrees of freedom associated to the left-handed modes are left unspecified. For such degrees of freedom, that correspond to a particular combination of the dreibein and the spin connection, one can further impose the torsionless condition consistently, obtaining in this way a theory with no local degrees of freedom whose asymptotic isometry group is generated by a single copy of the Virasoro algebra with central charge $c_{R}=3l/G$. This is reminiscent of what happens in Topologically Massive Gravity (TMG)\ at the chiral point. Nevertheless, it is worth to distinguish between the two constructions; we will comment on this distinction and on the analogies in section 5.

We begin in section 2 by reviewing chiral gravity. In section 3, we review the Mielke-Baekler theory of three-dimensional gravity, which includes torsion. In section 4 we review the calculation of the central charges for the theory with torsion, and we observe that a special point at which one of the central charges vanishes exists. In section 5 we discuss a prescription to approach the point of the
space of parameters at which Mielke-Baekler theory exhibits degeneracy, and we analyze the canonical structure of the theory. Section 6 contains the conclusions.

\section{Chiral gravity}

\subsection{Topologically massive gravity}

Let us start by discussing TMG \cite{Deser1982,Deser1982a}, which we review here within the context of \cite{Li2008}. The action of the theory, written in the first order formalism, reads
\begin{eqnarray}
S_{\text{TMG}} &=&\frac{1}{16\pi G}\int_{\Sigma _{3}}\varepsilon _{abc}\ R^{ab}\wedge e^{c}+\frac{\Lambda }{16\pi G}\int_{\Sigma _{3}}\varepsilon_{abc}\ e^{a}\wedge e^{b}\wedge e^{c}+  \nonumber \\
&&\frac{1}{32\pi G\mu }\int_{\Sigma _{3}}\varepsilon _{abc}(\omega^{ab}\wedge d\omega ^{c}+\frac{1}{3}\omega ^{a}\wedge \omega ^{b}\wedge \omega ^{c})+\frac{1}{32\pi G\mu }\int_{\Sigma _{3}}\lambda _{a}T^{a} ~,
\label{sTMG}
\end{eqnarray}
where the torsion 2-form \thinspace $T^{a}=\frac{1}{2}T_{\mu \nu }^{a}\ dx^{\mu }\wedge dx^{\nu }$ is defined by 
\[
T^{a}=de^{a}+\omega ^{ab}\wedge e_{b} ~,
\]
while the Riemannian curvature 2-form $R^{ab}=\frac{1}{2}R_{\ \mu \nu}^{ab}\ dx^{\mu }\wedge dx^{\nu }$ is defined by
\[
R^{ab}=d\omega ^{ab}+\omega _{\ c}^{a}\wedge \omega ^{cb} ~,
\]
with the dreibein 1-form $e^{a}=e_{\mu }^{a}dx^{\mu }$ and the spin connection 1-form $\omega ^{ab}=\omega _{\mu }^{ab}dx^{\mu }$. The convention adopted here is the standard one, according to which greek indices $\mu ,\nu ,\gamma ,...$ refer to spacetime coordinates while latin indices $a,b,c,...$ refer to coordinates in the tangent bundle; so we have $\omega ^{a}=\eta ^{ab}\omega _{b}$, $e^{a}=\eta ^{ab}e_{b}$, and the dual quantities like $\omega ^{a}=\varepsilon ^{abc}\omega _{bc}$, $R^{a}=\varepsilon _{\ bc}^{a}R^{bc}$, etc.

The first two terms in the gravitational action (\ref{sTMG}) correspond to the Einstein-Hilbert and the cosmological terms, with Newton constant $G$ and cosmological constant $\Lambda =-l^{-2}$ . The third contribution in (\ref{sTMG}) is the so-called 'exotic' gravitational Chern-Simons term, which
is purely made of the spin connection $\omega _{\mu }^{ab}$. Also, there is a fourth term in the action, which includes the torsion and a Lagrange multiplier $\lambda _{a}$. The Lagrange multiplier is actually a vector-valued 1-form $\lambda ^{a}=\lambda _{\mu }^{a}\ dx^{\mu}$, whose inclusion in the action implements the constraint of vanishing torsion $T^{a}=0$. The theory has a mass scale $\mu $, which turns out to be the mass of the gravitons of the theory \cite{Deser1982,Deser1982a}.

The equations of motion coming from the action above are
\begin{eqnarray}
\varepsilon _{abc}\left( R^{bc}+\frac{1}{l^{2}}e^{b}\wedge e^{c}\right) - \frac{1}{\mu }D\lambda _{a} &=&0 ~, \label{Eom1} \\ [0.5em]
R^{ab}+\frac{1}{2}\left( \lambda ^{a}\wedge e^{b}-e^{a}\wedge \lambda^{b}\right) +\mu\,\varepsilon ^{abc}T_{c} &=&0 ~, \label{Eom2} \\ [0.8em]
T^{a} &=&0 ~,
\label{Eom3}
\end{eqnarray}
where the 2-form $D\lambda _{a}=d\lambda _{a}+\omega _{ab}\wedge \lambda^{b} $ is the covariant derivative of the Lagrange multiplier. These equations correspond to varying action (\ref{sTMG}) with respect to the dreibein, the spin connection, and the Lagrange multiplier, respectively. Notice that this is different from what happens in three-dimensional general relativity, where the equation of motion $T^{a}=0$ comes from varying the Einstein-Hilbert action with respect to the spin connection instead. For a concise review of TMG\ in the first order formalism we refer to the recent papers \cite{Grumiller2008a,Miskovic2009a}.

Using equation (\ref{Eom3}) above one may write the set of field equations as follows
\begin{eqnarray}
\varepsilon _{abc}\left( R^{bc}+\frac{1}{l^{2}}e^{b}\wedge e^{c}\right) - \frac{1}{\mu }D\lambda _{a} &=&0 ~, \label{EOM1} \\ [0.7em]
R^{ab}+\frac{1}{2}\left( \lambda ^{a}\wedge e^{b}-e^{a}\wedge \lambda^{b}\right)  &=&0 ~, \label{EOM2}
\end{eqnarray}
and from (\ref{EOM2}), which is an algebraic equation, one solves for $\lambda _{\mu }^{a}$ and replace it back in (\ref{EOM2}) to obtain the Cotton tensor made of $D\lambda _{a}$. This defines TMG in the form we know it \cite{Deser1982,Deser1982a}. The theory, thus, corresponds to a dynamical theory with equations of motion of third order that includes general relativity as a particular sector. In fact, it is well known that all classical solutions to three-dimensional general relativity solve the equations of TMG as well; this is basically because the Cotton tensor vanishes if (and only if) the metric is conformally flat.

\subsection{Asymptotically AdS solutions}

Here, we are concerned with asymptotically AdS$_{3}$ geometries. Written in a convenient system of coordinates, (a patch of) AdS$_{3}$ space reads
\[
ds_{\text{AdS}}^{2}=-\left( \frac{r^{2}}{l^{2}}+1\right) dt^{2}+\left( \frac{r^{2}}{l^{2}}+1\right) ^{-1}dr^{2}+r^{2}d\phi ^{2} ~,
\]
where $l$ is the 'radius' of AdS$_{3}$ space. In this system of coordinates the asymptotically AdS$_{3}$ boundary conditions take the form
\begin{eqnarray}
g_{tt} \simeq -\displaystyle\frac{r^{2}}{l^{2}}+\mathcal{O}(1) ~, \qquad
& g_{tr} \simeq \mathcal{O}(1/r^{3}) ~, \qquad
& g_{t\phi} \simeq \mathcal{O}(1/r^{3}) ~,\qquad \label{BH1} \\ [0.7em]
g_{rr} \simeq \displaystyle\frac{l^{2}}{r^{2}}+\mathcal{O}(1/r^{4}) ~, \qquad
& g_{r\phi} \simeq \mathcal{O}(1/r^{3}) ~, \qquad
& g_{\phi\phi} \simeq r^{2}+\mathcal{O}(1) ~.\qquad  \label{BH2}
\end{eqnarray}
This is the set of boundary conditions introduced by Brown and Henneaux in \cite{Brown1986b}.

The most important member of the set of asymptotically AdS$_{3}$ solutions of general relativity is the Ba\~{n}ados-Teitelboim-Zanelli black hole (BTZ), whose metric is given by \cite{Banados1992,Banados1993}
\begin{eqnarray}
ds_{\text{BTZ}}^{2} &=& -\left( \frac{r^{2}}{l^{2}}-8GM\right) dt^{2}+\left( \frac{r^{2}}{l^{2}}-8GM+\frac{16G^{2}M^{2}l^{2}}{r^{2}}\right)^{-1}dr^{2} \nonumber \\ [0.7em]
& & \qquad+\,r^{2}d\phi ^{2}+8GJ\ d\phi\,dt ~.  \label{BTZ}
\end{eqnarray}
Indeed, taking a glance at the asymptotic conditions above it is evident that this metric is part of the set of solutions considered in the analysis of \cite{Brown1986b}.

BTZ black hole (\ref{BTZ}), thought of as a solution to TMG, has mass and angular momentum given by
\begin{equation}
\mathcal{M}=M+\frac{1}{\mu l^{2}}J ~, \qquad \mathcal{J}=J+\frac{1}{\mu }M ~,
\label{MJ}
\end{equation}
respectively. These reduce to the ADM\ values of general relativity when $1/\mu =0$.

Asymptotic conditions (\ref{BH1})-(\ref{BH2}) are easily expressed in the first order formalism: The dreibein $e_{\mu }^{a}$ is, up to a local Lorentz transformation, defined in terms of the metric by $g_{\mu \nu }=e_{\mu}^{a}e_{\nu }^{b}\eta _{ab}$. Then, Brown-Henneaux boundary conditions (%
\ref{BH1})-(\ref{BH2}) for the components $e_{\mu }^{a}$ read
\begin{eqnarray*}
e_{t}^{0} \simeq \displaystyle\frac{r}{l}+\mathcal{O}(1/r) ~, \qquad
& e_{r}^{0} \simeq \mathcal{O}(1/r^{4}) ~, \qquad
& e_{\phi }^{0} \simeq \mathcal{O}(1/r) ~, \\ [0.7em]
e_{t}^{1} \simeq \mathcal{O}(1/r^{2}) ~, \qquad
& e_{r}^{1} \simeq \displaystyle\frac{l}{r}+\mathcal{O}(1/r^{3}) ~, \qquad
& e_{\phi }^{1}\simeq \mathcal{O}(1/r^{2}) ~, \\ [0.7em]
e_{t}^{2} \simeq \mathcal{O}(1/r) ~, \qquad
& e_{r}^{2}\simeq \mathcal{O}(1/r^{4}) ~, \qquad
& e_{\phi }^{2}\simeq r+\mathcal{O}(1/r) ~.
\end{eqnarray*}
And from the vanishing torsion constraint, $T^{a}=de^{a}+\omega_{\ b}^{a}\wedge e^{b}=0$, one obtains the falling-off conditions for the components $\omega _{\mu }^{a}$ of the spin connection as well.

\subsection{The chiral gravity conjecture}

The algebra of conserved charges associated to the asymptotic isometry group of TMG\ in AdS$_{3}$ is generated by two copies of the Virasoro algebra, as it happens in the case of general relativity \cite{Brown1986b}. The central charges associated to each of these Virasoro algebras can be computed by several methods, and turn out to be
\begin{equation}
c_{L}=\frac{3l}{2G}\left( 1-\frac{1}{\mu l}\right) ~,\quad \quad \quad \quad
c_{R}=\frac{3l}{2G}\left( 1+\frac{1}{\mu l}\right) ~,  \label{CTMG}
\end{equation}
which reproduce the result of Brown and Henneaux for general relativity in the case $1/\mu =0$, namely $c_{L}=c_{R}=3l/(2G)$.

One of the observations made in \cite{Li2008}, and which motivated the whole idea of a chiral theory of gravity in three dimensions, is that at the special point of the space of parameters where $\mu l=1$, the left-handed central charge $c_{L}$ vanishes. Besides, according to (\ref{MJ}), the mass and the angular momentum of a generic BTZ\ black hole at $\mu l=1$ obey the relation $\mathcal{M}l=\mathcal{J}$, no matter the values that the parameters $M$ and $J$ take. In particular, this implies that all the BTZ\ black holes (\ref{BTZ}) are extremal states. It also implies a plethora of solutions with vanishing conserved charges, which correspond to those BTZ\ metrics (\ref{BTZ}) with parameters $Ml=-J$. Besides, it is possible to see that in the limit $\mu l\rightarrow 1$ the massive graviton of TMG tends to one of the modes of Einstein gravity, which is pure gauge. All these suggestive facts were gathered as pieces of evidence, and led the authors of \cite{Li2008} to conjecture that at the point $\mu l=1$ TMG\ about AdS$_{3}$ space becomes a bulk theory with no local degrees of freedom that would be dual to a chiral conformal field theory with $c_{R}=3l/G$; see \cite{Carlip2008,Carlip2009,Li2008a,Giribet2008b,Carlip2008c,Carlip2009c,Henneaux2009,Henneaux2010,Gaberdiel2010,Cunliff2010} for discussions.

A rapid way to notice that TMG exhibits special features at $\mu l=1$ is to consider \textit{pp}-waves in AdS$_{3}$ \cite{Ayon-Beato2006,Ayon-Beato2009}. Consider the exact solution
\begin{equation}
ds^{2}=-\frac{r^{2}}{l^{2}}F(u,r)du^{2}-2\frac{r^{2}}{l^{2}}du\,dv+\frac{l^{2}}{r^{2}}dr^{2} ~,  \label{waves}
\end{equation}
which corresponds to a non-linear solution of the equations of motion (\ref{EOM1})-(\ref{EOM2}) whose physical interpretation is that of a \textit{pp}-wave sailing the AdS$_{3}$ spacetime. AdS$_{3}$ spacetime written in Poincar\'{e} coordinates corresponds $F(u,r)=0$, and one can identify the coordinates as $u=t-\phi $ and $v=t+\phi $, so that the front of the wave corresponds to the surfaces $u=v=$ const. The function $F(u,r)$ gives the profile of the wave, which takes the form $F(u,r)=\left( r/l\right) ^{\mu l -1}f(u)$. This function satisfies the scalar wave equation on AdS$_{3}$, namely $(\square -m_{\text{eff}}^{2})F(u,r)=0$, where $\square $ stands for the D'Alembert operator in AdS$_{3}$, and the effective mass $m_{\text{eff}}$ is given by $m_{\text{eff}}^{2}=\mu ^{2}(1-\mu ^{-2}l^{-2})$. That is, the profile function $F(u,r)$ behaves as a scalar mode of the space on which the non-linear wave solution is propagating. Then, one immediately notices that in the limit $\mu l\rightarrow 1$ such scalar mode becomes massless. One can also verify that non-linear solutions (\ref{waves}) develop a logarithmic falling-off behavior at the boundary; namely, solutions like $\sim \log (r/l)$ arise at $\mu l=1$.

The consistency of chiral gravity was matter of intense discussion recently \cite{Carlip2008,Carlip2009,Li2008a,Giribet2008b,Carlip2008c,Carlip2009c,Henneaux2009,Henneaux2010,Gaberdiel2010,Cunliff2010}, and complete consensus has not yet been reached. Nevertheless, the idea is very promising and, besides, it gave rise to interesting results as byproduct. The discussion about chiral gravity was mainly about its spectrum, as it is crucial to establish the consistency of the whole construction. Consequently, the field content of the theory was analyzed in extent, both at the linearized level and at the level of exact solutions. On the one hand, in what regards to linearized solutions, the discussion is summarized in \cite{Maloney2010}, where it was understood that at $\mu l=1$ two different set of boundary conditions are admissible: the one proposed by Brown and Henneaux in \cite{Brown1986b}, and the weakened version proposed by Grumiller and Johansson in \cite{Grumiller2008,Grumiller2009,Ertl2009}; and depending on which of these asymptotics is chosen, the resulting theory happens to exhibit different properties. In particular, the boundary conditions proposed in \cite{Grumiller2008,Grumiller2009,Ertl2009} permit asymptotic behaviors like
\begin{eqnarray}
& g_{tt} \simeq -\displaystyle\frac{r^{2}}{l^{2}}+\mathcal{O}(\log (r)) ~, \qquad
& g_{\phi t}\simeq \mathcal{O}(\log (r)) ~, \label{q1} \\ [0.7em]
& g_{rr} \simeq \displaystyle\frac{l^{2}}{r^{2}}+\mathcal{O}(r^{-4}) ~, \qquad
& g_{\phi \phi}\simeq r^{2}+\mathcal{O}(\log (r)) ~, \label{q2}
\end{eqnarray}
which are certainly weaker than (\ref{BH1})-(\ref{BH2}). If these boundary conditions are chosen, the bulk theory has ghosts \cite{Grumiller2008,Grumiller2009,Ertl2009} and the boundary CFT renders non-unitary\footnote{G.G. thanks M. Kleban and M. Porrati for illuminating discussions and collaboration on this question.} \cite{Grumiller2010d,Grumiller2011}.

On the other hand, in what regards to the analysis of the exact solutions, it was shown in \cite{Garbarz2009} that solutions satisfying the weakened asymptotics (\ref{q1})-(\ref{q2}) without satisfying (\ref{BH1})-(\ref{BH2}) do exist. The existence of such solutions situates the discussion of boundary conditions beyond the linearized analysis. One such a solution is given by 
\begin{equation}
ds^{2}=-\frac{r^{2}}{l^{2}}dt^{2}+\frac{l^{2}}{r^{2}}dr^{2}+r^{2}d\phi
^{2}+k\log \left( r^{2}/r_{0}^{2}\right) (dt-ld\phi )^{2} ~, \label{Log}
\end{equation}
which corresponds to deforming a special case of the BTZ\ geometry (\ref{BTZ}) by adding a logarithmic piece, where $k$ is an integration constant associated to the mass and angular momentum of the solution; more precisely $\mathcal{M}l=\mathcal{J}\sim k$ \cite{Garbarz2009,Miskovic2009a}. Then, the next
question about exact solutions that one might feel tempted to ask is whether by imposing strong boundary conditions (\ref{BH1})-(\ref{BH2}), instead of (\ref{q1})-(\ref{q2}), the classical sector of TMG\ at $\mu l=1$ coincides with that of Einstein gravity or not. If the classical sectors of both theories were the same, then the theory would not admit physical local degrees of freedom. This is equivalent to asking whether equations (\ref{EOM1})-(\ref{EOM2}) supplemented with boundary conditions (\ref{BH1})-(\ref{BH2}) imply that the Cotton tensor vanishes necessarily. This question was answered by the negative in \cite{Compere2010} where an exact solution to TMG at $\mu l=1$ obeying Brown-Henneaux asymptotic without being Einstein manifolds were found. Such a solution is given by
\begin{equation}
ds^{2}=-\frac{r^{2}}{l^{2}}dt^{2}+\frac{l^{2}}{r^{2}}dr^{2}+r^{2}d\phi^{2}+\left( \frac{\gamma t}{l^{2}}-\frac{\gamma ^{2}l^{2}}{96r^{4}}\right) (dt+ld\phi )^{2} ~, \label{belgas}
\end{equation}
where $\gamma $ is a parameter. This is a time-dependent solution of the theory at $\mu l=1$, and also corresponds to a deformation of a special case of (\ref{BTZ}). Metric (\ref{belgas}) solves (\ref{EOM1})-(\ref{EOM2}) having non-vanishing Cotton tensor. Solutions like (\ref{belgas}), however, seem to carry vanishing conserved charges, $\mathcal{M}=\mathcal{J}=0$, so they would not contribute substantially to the partition function. Then, the next question to be asked is whether non-Einstein solutions to TMG at $\mu l=1$ with Brown-Henneaux boundary conditions and finite mass actually exist. To the best of our knowledge, this remains an open question.

\section{Adding torsion}

\subsection{Mielke-Baekler theory of gravity}

A different construction of a chiral theory in AdS$_{3}$ is possible. This consists of considering a special case of three-dimensional Chern-Simons gravity including torsion \cite{Mardones1991}, also known as Mielke-Baekler theory \cite{Mielke1991}. The action of the theory can be written as follows
\begin{equation}
S_{\text{MB}}=\frac{1}{16\pi G}S_{1}+\frac{\Lambda }{16\pi G}S_{2}+\frac{1}{16\pi G\mu }S_{3}+\frac{m}{16\pi G}S_{4} ~, \label{SMB}
\end{equation}
where the four terms are
\begin{equation}
S_{1}=2\int_{\Sigma _{3}}e_{a}\wedge R^{a} ~, \qquad S_{2}=-\frac{1}{3} \int_{\Sigma _{3}}\varepsilon _{abc}\,e^{a}\wedge e^{b}\wedge e^{c} ~,
\end{equation}
\begin{equation}
S_{3}=\int_{\Sigma _{3}}(\omega _{a}\wedge d\omega ^{a}+\frac{1}{3} \varepsilon _{abc}\,\omega ^{a}\wedge \omega ^{b}\wedge \omega ^{c}) ~,\ \ \ S_{4}=\int_{\Sigma _{3}}e_{a}\wedge T^{a} ~.
\end{equation}
Here, again, we see that in addition to the Einstein-Hilbert action, $S_{1}$, and the cosmological constant term, $S_{2}$, we have the exotic Chern-Simons gravitational term, $S_{3}$, together with the term $S_{4}$ that involves the torsion explicitly. In fact, there are actually two stages at which one introduces torsion here: first, this is done by treating the dreibein, $e^{a}$, and the spin connection, $\omega ^{a}$, as independent fields, following in this way the standard formulation \`{a} la Einstein-Cartan. In the case of general relativity, the Palatini formulation teaches us that considering $e^{a}$ and $\omega ^{a}$ as independent variables does not introduce any substantial difference for the classical theory, as the Einstein equations are recovered by varying the Einstein-Hilbert action with respect to $e^{a}$, while the vanishing torsion constraint follows from varying the action with respect to $\omega ^{a}$. However, when the exotic gravitational Chern-Simons term is present in the action, the fact of treating $e^{a}$ and 
$\omega ^{a}$ as disconnected geometrical entities does make an important difference.

A second stage at which one introduces torsion in the theory is by adding the term $S_{4}$ when writing the action. Such term includes the torsion explicitly, and, in contrast to (\ref{sTMG}), it does not involve a Lagrange multiplier that fixes the torsion to zero but it couples the torsion to the dreibein directly. The term $S_{4}$ is dubbed 'translational Chern-Simons term'\ and, as it happens with the exotic Chern-Simons term $S_{3}$, it can be also associated to a topological invariant in four dimensions: While $S_{3}$ is thought of as the term whose (dimensionally extended) exterior derivative gives the Pontryagin 4-form density $R^{ab}\wedge R_{ab}$ in four dimensions, the exterior derivative of the translational term $S_{4}$ gives the Nieh-Yan 4-form density $T^{a}\wedge T_{a}-e^{a}\wedge e^{b}\wedge R_{ab}$, \cite{Nieh1982,Chandia1997}. In this sense, all the terms involved in the action (\ref{SMB}) are of the same sort \cite{Zanelli2005}.

The equations of motion coming from (\ref{SMB}) are
\begin{eqnarray}
R^{a}-\frac{\Lambda }{2}\ \varepsilon _{\,\,\,bc}^{\,a}\,e^{b}\wedge e^{c}+m\ T^{a} &=&0 ~, \label{eom2} \\ [0.7em]
T^{a}+\frac{1}{\mu }\ R^{a}+\frac{m}{2}\ \varepsilon_{\,\,\,bc}^{\,a}\,e^{b}\wedge e^{c} &=&0 ~. \label{2eom}
\end{eqnarray}
The first one comes from varying $S_{\text{MB}}$ with respect to the dreibein, while the second one comes from varying it with respect to the spin connection. One actually sees that in the case $m=1/\mu =0$ the theory agrees with Einstein's gravity, for which $R^{ab}\sim e^{a}\wedge e^{b}$ and $T^{a}=0$. In the special case $m=\mu $ with $\Lambda =-m^{2}$, the two equations of motion (\ref{eom2}) and (\ref{2eom}) coincide and the theory exhibits a degeneracy. We will analyze this special case in section 5. In the generic case, the theory has four coupling constants, which provide three
dimensionless ratios, and the four characteristic length scales $G$, $\sqrt{\Lambda}$, $\mu^{-1}$, and $m^{-1}$.

The two equations of motion (\ref{eom2})-(\ref{2eom}) are independent equations provided $m\neq \mu $; so let us consider such case first. Arranging these equations, one finds
\begin{equation}
R^{a}=\frac{\mu }{2}\frac{\Lambda +m^{2}}{\mu -m}\,\varepsilon_{\,\,\,bc}^{\,a}\,e^{b}\wedge e^{c} ~, \label{Rcte}
\end{equation}
\begin{equation}
T^{a}=\frac{1}{2}\frac{m\mu +\Lambda }{m-\mu }\,\varepsilon_{\,\,\,bc}^{\,a}\,e^{b}\wedge e^{c} ~.  \label{Tcte}
\end{equation}
These equations, provided $m\neq \mu$, express the fact that the solutions of the theory are constant curvature and constant torsion. From (\ref{Rcte})-(\ref{Tcte}) one immediately identifies two special cases. When $m\mu = -\Lambda$ ($m\neq \mu$), equation (\ref{Tcte}) implies that torsion vanishes, and thus (\ref{Rcte}) becomes the Einstein equations. A second special case is $m^{2}=-\Lambda$ ($m\neq \mu$), where it is the spacetime curvature what vanishes; this is usually called the 'teleparallel theory'.

Even though equation (\ref{Rcte}) implies that the solutions of the theory have to be of constant curvature, the space has torsion, so that the affine connection is not necessarily a Levi-Civita connection. Then, seeing whether the solutions of the theory actually\ correspond to Einstein manifolds or not requires a little bit more of analysis: To actually see this, it is convenient to write the spin connection $\omega ^{a}$ as the sum of a torsionless contribution $\tilde{\omega}^{a}$ and the contorsion $\Delta \omega ^{a}$; namely 
\begin{equation}
\omega ^{a}=\tilde{\omega}^{a}+\Delta \omega ^{a} ~, \label{wdec}
\end{equation}
where $\tilde{\omega}^{a}$ is indeed the Levi-Civita connection. Then, from (\ref{Tcte}) one obtains 
\begin{equation}
\Delta \omega ^{a}=\frac{1}{2}\frac{m\mu +\Lambda }{m-\mu }\ e^{a} ~,
\label{kcond}
\end{equation}
and from (\ref{Rcte}) one finally gets
\begin{equation}
\tilde{R}^{ab}=d\tilde{\omega}^{ab}+\tilde{\omega}_{c}^{a}\wedge \tilde{\omega}^{cb}=-\frac{1}{2l^{2}}\,\ e^{a}\wedge e^{b} ~, \label{R0}
\end{equation}
which expresses that solutions are indeed Einstein manifolds, where the effective cosmological constant is given by
\begin{equation}
l^{-2}=\frac{1}{4}\left( \frac{m\mu +\Lambda }{m-\mu }\right) ^{2}+\frac{\Lambda \mu +m^{2}\mu }{m-\mu} ~. \label{effectivecons}
\end{equation}
In the case $m=1/\mu =0$ one finds $l^{-2}=-\Lambda $.

\subsection{Black holes and torsion}

Mielke-Baekler theory admits asymptotically AdS$_{3}$ black holes as exact solutions. In fact, it can be seen that equations of motion (\ref{eom2})-(\ref{2eom}) are satisfied by the BTZ\ metric (\ref{BTZ}), provided the space also presents torsion \cite{Garcia2003}. The presence of non-vanishing torsion, however, does not represent an actual 'hair' since the strength of $T^{a}$ is fixed by (\ref{Tcte}) and so there is no additional parameter to characterize the geometry. Then, the only two parameters of the black hole solutions are still $M$ and $J$, and for the Mielke-Baekler theory, the mass and angular momentum of the black hole being related to the coupling constants in the following way
\begin{equation}
\mathcal{M}=M\left( 1+\frac{1}{2}\frac{m\mu +\Lambda }{m\mu -\mu ^{2}}+\frac{J}{Ml^{2}\mu }\right) ~, \qquad \mathcal{J}=J\left( 1+\frac{1}{2}\frac{m\mu + \Lambda }{m\mu -\mu ^{2}}+\frac{M}{J\mu }\right) ~. \label{MJ2}
\end{equation}
The ADM\ values of general relativity are recovered in the case $m=1/\mu =0$.

Black hole thermodynamics is also affected by the presence of torsion. The entropy of the BTZ\ black holes in Mielke-Baekler theory can be computed, and is given by
\begin{equation}
S_{\text{BH}}=\frac{\pi r_{+}}{2G}\left( 1+\frac{1}{2}\frac{m\mu +\Lambda}{m\mu -\mu ^{2}}-\frac{1}{\mu l}\frac{r_{-}}{r_{+}}\right) ~, \label{entropy1}
\end{equation}
where $r_{+}$ and $r_{-}$ are the horizons of the black hole, namely
\begin{equation}
r_{\pm }^{2}=4l^{2}GM\left( 1\pm \sqrt{1-\frac{J^{2}}{M^{2}l^{2}}}\right) ~.
\label{horizontes}
\end{equation}
While the first term in (\ref{entropy1}) reproduces the Bekenstein-Hawking area law, contributions proportional to $1/\mu $ give deviations from the result of general relativity. It will be discussed below how the black hole entropy (\ref{entropy1}) is recovered from CFT methods through holography.

\section{Central charges}

\subsection{Chern-Simons formulation and Hamiltonian reduction}

In this section, we focus on the computation of the central charges corresponding to the asymptotic algebra. Seeing from the holographic point of view, these central charges turn out to be those of the dual conformal field theory. To calculate these central charges it is convenient to discuss first the Chern-Simons formulation of the theory (\ref{SMB}). In fact, Mielke-Baekler theory admits to be expressed as a sum of two Chern-Simons actions \cite{Blagojevic2003a,Blagojevic2003,Blagojevic2003b,Klemm2008,Cacciatori2006},
\begin{equation}
S_{\text{CS}}=k\int \text{tr}\left( A\wedge dA+\frac{2}{3}A\wedge A\wedge A\right) -\hat{k}\int \text{tr}\left( \hat{A}\wedge d\hat{A}+\frac{2}{3}\hat{A}\wedge \hat{A}\wedge \hat{A}\right) ~, \label{SCS}
\end{equation}
where, for the case of the theory with negative cosmological constant $\Lambda =-l^{-2}$, the corresponding $SL(2,\mathbb{R})$ connections are given by 
\begin{equation}
A^{a}=\omega ^{a}+\lambda \ e^{a} ~,\quad \quad \quad \hat{A}^{a}=\omega ^{a}+\hat{\lambda}\ e^{a} ~,  \label{cuatropuntouno}
\end{equation}
with coefficients
\begin{equation}
\lambda =-\frac{1}{2}\frac{m\mu +\Lambda }{m-\mu }+\frac{1}{l} ~, \qquad \hat{\lambda}=-\frac{1}{2}\frac{m\mu +\Lambda }{m-\mu }-\frac{1}{l} ~;
\end{equation}
whereas the coupling constants read
\begin{equation}
k=\frac{l}{32\pi G}\left( 1+\frac{1}{\mu l}+\frac{1}{2}\frac{m\mu +\Lambda }{m\mu -\mu ^{2}}\right) ~, \qquad \hat{k}=\frac{l}{32\pi G}\left( 1-\frac{1}{\mu l}+\frac{1}{2}\frac{m\mu +\Lambda }{m\mu -\mu ^{2}}\right) ~.
\end{equation}
The index $a$ in (\ref{cuatropuntouno}) now is playing the r\^{o}le of group index, to be contracted with the $3+3$ generators of the $sl(2)\oplus sl(2)$ algebra. This is analogous to the standard Chern-Simons realization of three-dimensional gravity and, in fact, in the case $m=0$ the realization of \cite{Achucarro1986,Witten1988a} is recovered. In this formulation, the equations of motion of the theory read
\begin{equation}
F=0,\qquad \widehat{F}=0 ~,
\end{equation}
where $F$ and $\widehat{F}$ are the field strength corresponding to the gauge fields $A$ and $\hat{A}$ respectively.

Having the theory written in its form (\ref{SCS}), one can compute the central charges by following the procedure originally introduced in \cite{Coussaert1995,Henneaux2000}. This amounts to implementing AdS$_{3}$ asymptotic boundary conditions at the level of the Chern-Simons actions, by reducing them first to two chiral Wess-Zumino-Witten (WZW) actions, and then using the asymptotic conditions again to reduce some degrees of freedom of the latter. This eventually gives the central charges of the boundary two-dimensional conformal field theory through the Hamiltonian reduction of the WZW theory, as in \cite{Coussaert1995,Henneaux2000}. Nevertheless, despite the analysis here is very similar to that of three-dimensional Einstein gravity, it is worth noticing that, in contrast to the case where no exotic Chern-Simons term is included, the full action is not exactly the difference of two chiral WZW actions with the same level $k=\hat{k}$. The exotic term actually unbalances the two chiral contributions. In turn, Hamiltonian reduction must be performed in each piece separately.

Consistent set of AdS$_{3}$ boundary conditions for the theory with torsion are those proposed in \cite{Blagojevic2003a,Blagojevic2003,Blagojevic2003b}
\begin{eqnarray}
e_{t}^{0} \simeq \displaystyle\frac{r}{l}+\mathcal{O}(1/r) ~, \qquad
& e_{r}^{0} \simeq \mathcal{O}(1/r^{4}) ~, \qquad
& e_{\phi }^{0} \simeq \mathcal{O}(1/r) ~, \nonumber \\ [0.7em]
e_{t}^{1} \simeq \mathcal{O}(1/r^{2}) ~, \qquad
& e_{r}^{1} \simeq \displaystyle\frac{l}{r}+\mathcal{O}(1/r^{3}) ~, \qquad
& e_{\phi }^{1} \simeq \mathcal{O}(1/r^{2}) ~, \label{t1} \\ [0.7em]
e_{t}^{2} \simeq \mathcal{O}(1/r) ~, \qquad
& e_{r}^{2} \simeq \mathcal{O}(1/r^{4}) ~, \qquad
& e_{\phi }^{2} \simeq r+\mathcal{O}(1/r) ~. \nonumber
\end{eqnarray}
From equation (\ref{Tcte}), one obtains the asymptotic behavior for the components of the spin connection; namely
\begin{eqnarray}
\omega_{t}^{0} \simeq \displaystyle\frac{ar}{2l}+\mathcal{O}(1) ~, \qquad
& \omega_{r}^{0} \simeq \mathcal{O}(1/r^{4}) ~, \qquad
& \omega_{\phi }^{0} \simeq -\displaystyle\frac{r}{l}+\mathcal{O}(1) ~, \nonumber \\ [0.7em]
\omega_{t}^{1} \simeq \mathcal{O}(1/r^{2}) ~, \qquad
& \omega_{r}^{1} \simeq \displaystyle\frac{al}{2r}+\mathcal{O}(1/r^{3}) ~, \qquad
& \omega_{\phi }^{1} \simeq \mathcal{O}(1/r^{2}) ~, \label{t2} \\ [0.7em]
\omega_{t}^{2} \simeq -\displaystyle\frac{r}{l^{2}}+\mathcal{O}(1/r) ~, \qquad
& \omega_{r}^{2} \simeq \mathcal{O}(1/r^{4}) ~, \qquad
& \omega_{\phi }^{2} \simeq \displaystyle\frac{ar}{2}+\mathcal{O}(1/r) ~, \nonumber
\end{eqnarray}
where $a=(m\mu +\Lambda )/(m-\mu )$.

Then, following the procedure developed in \cite{Coussaert1995,Henneaux2000}, one verifies that implementing some of the asymptotic conditions (\ref{t1})-(\ref{t2}) amounts to define a boundary action, consisting of two copies of the chiral WZW model (see \cite{Coussaert1995,Henneaux2000} for details, and see also \cite{Nitti2004} for a very nice discussion).\ The WZW theory has $SL(2,\mathbb{R})_{k}\times SL(2,\mathbb{R})_{\hat{k}}$ affine Kac-Moody symmetry, which is generated by the currents
\[
J^{i}(z)=\sum\limits_{n}J_{n}^{i}\ z^{-n-1} ~, \quad \bar{J}^{i}(z)=\sum\limits_{n}\bar{J}_{n}^{i}\ \bar{z}^{-n-1} ~, \quad \quad i=1,2,3 ~,
\]
with the boundary variables $z=t+i\phi $, $\bar{z}=t-i\phi $. The modes obey the Kac-Moody current algebra
\[
\lbrack J_{m}^{+},J_{n}^{-}]=-2J_{n+m}^{3}-\frac{k}{2}n\,\delta_{m+n,0} ~, \qquad
\lbrack J_{m}^{3},J_{n}^{\pm }]=\pm J_{n+m}^{\pm} ~, \qquad
\lbrack J_{m}^{3},J_{n}^{3}]=\frac{k}{2}n\delta_{m+n,0} ~,
\]
with $J_{n}^{\pm}=J_{n}^{1}\pm iJ_{n}^{2}$, where $k$ is a central element; analogously for the anti-holomorphic counterpart $\bar{J}_{n}^{i}$ with $\hat{k}$. Then, Sugawara construction gives the Virasoro generators in terms of the Kac-Moody generators; namely
\begin{equation}
L_{m}=\frac{h_{ij}}{k-2}\sum\limits_{n}J_{m-n}^{i}\,J_{n}^{j} ~, \qquad
\bar{L}_{m}=\frac{h_{ij}}{k-2}\sum\limits_{n}\bar{J}_{m-n}^{i}\,\bar{J}_{n}^{j} ~,
\end{equation}
where $h_{ij}$ is the Cartan-Killing bilinear form of $SL(2,\mathbb{R})$ and the $-2$ in the denominator stands for the Coxeter number of $SL(2,\mathbb{R})$. Then, we have the stress-tensor
\begin{equation}
T(z)=\sum\limits_{n}L_{n}\ z^{-n-2} ~, \qquad
\overline{T}(\bar{z})=\sum\limits_{n}\bar{L}_{n}\ \bar{z}^{-n-2} ~,
\end{equation}
whose modes realize the Virasoro algebra
\begin{equation}
\lbrack L_{m},L_{n}]=(m-n)\,L_{m+n}+\frac{k}{4(k-2)}m^{2}(m^{2}-1)\,\delta_{m+n,0} ~,
\end{equation}
that gives the central charges $c=3k/(k-2),$ and analogously for the anti-holomorphic counterpart replacing $L_{n}$ by $\bar{L}_{n}$ and $k$ by $\hat{k}$, yielding $\widehat{c}=3\hat{k}/(\hat{k}-2)$. These are not yet the central charges of the boundary CFT as it still remains to impose some of the boundary conditions (\ref{t1})-(\ref{t2}). It is possible to verify that implementing the whole set of asymptotic boundary conditions (\ref{t1})-(\ref{t2}) amounts to fixing the constraints $J^{+}(z)\equiv k$ and $\bar{J}^{+}(\bar{z})\equiv \hat{k}$. This condition requires an improvement of the stress-tensor of the sort $T(z)\rightarrow T(z)+\partial J^{3}(z)$, as it demands
the current $J^{+}(z)$ to be a dimension zero field. This is equivalent to shifting $L_{n}\rightarrow L_{n}-(n+1)J_{n}^{3}$, and the same for $\bar{L}_n$, which results in a shifting of the value of the central charges $c$ and $\widehat{c}$. The central charges now become $c_{R}=3k/(k-2)+6k$ and $c_{L}=3\hat{k}/(\hat{k}-2)+6\hat{k}$, and for large $k$, $\hat{k}$ one gets the standard result $c_{R}\simeq 6k$ and $c_{L}\simeq 6\hat{k}$. Then, one finds\footnote{A.G. and G.G. thank Mat\'{\i}as Leoni for discussions about this calculation in the case of having torsion.}
\begin{equation}
c_{L}=\frac{3l}{2G}\left( 1-\frac{1}{\mu l}+\frac{1}{2}\frac{m\mu +\Lambda}{m\mu -\mu ^{2}}\right) ~, \qquad c_{R}=\frac{3l}{2G}\left( 1+\frac{1}{\mu l}+\frac{1}{2}\frac{m\mu +\Lambda }{m\mu -\mu ^{2}}\right) ~, \label{cRcL}
\end{equation}
together with (\ref{effectivecons}). One rapidly verifies that this result agrees with the ones obtained in the literature \cite{Blagojevic2003a,Blagojevic2003,Blagojevic2003b,Klemm2008,Cacciatori2006}.

It is important to point out that, even in the case $m=0$, where the action of the theory only contains the Einstein-Hilbert and the cosmological terms, $S_{1}+S_{2}$, and the exotic Chern-Simons gravitational term, $S_{3}$, these values for the central charges do not coincide with those of TMG. This is because, as mentioned earlier, both theories differ not only because of the inclusion of $S_{4}$ in the action. In fact, if $m=0$, and taking (\ref{effectivecons}) into account, one finds $c_{L}=(3/2G\mu )(\sqrt{1+\mu^{2}l^{2}}-1)$, $c_{R}=(3/2G\mu )(\sqrt{1+\mu ^{2}l^{2}}+1)$, which
coincides with (\ref{CTMG}) only at first order in $\,1/\mu $. On the other hand, if $m\neq 0$ and $1/\mu =0$, the central charges above simply become $c_{L}=c_{R}=3l/(2G)$. This does not imply that the value of $m$ dissapears from the expressions since (\ref{effectivecons}) depends on $m$ and, thus, when $1/\mu=0$ the effective cosmological constant is given by $-l^{-2}=\Lambda +m^{2}$. This can be simply seen taking a glance at the equations of motion and noticing that replacing $1/\mu =0$ in (\ref{eom2})-(\ref{2eom}) makes the curvature to dissapear from (\ref{eom2}), while inducing at the same time a redefinition of the cosmological constant in (\ref{2eom}).

\subsection{Quantization conditions}

So, we have central charges (\ref{cRcL}). These are the central elements of the asymptotic AdS$_{3}$ isometry algebra, and from the AdS/CFT conjecture point of view these are the charges of the dual CFT. Modular invariance of such CFT demands $(c_{L}-c_{R})/24=(8G\mu)^{-1}\in \mathbb{Z}$, giving a
quantization condition for the parameters in the action. Besides, even before resorting to the dual CFT description, one may argue that the central charges have to be quantized. Indeed, quantization of the $SL(2,\mathbb{R})$ Chern-Simons coefficient imposes conditions on $c_{R}=6k$ and $c_{L}=6\hat{k}$ as well. For instance, already in the case $1/\mu =0$, one finds $(16G\sqrt{-\Lambda })^{-1}\in \mathbb{Z}$. This follows from topology arguments; see the nice discussion in \cite{Witten2007}.

As Witten pointed out also in \cite{Witten2007}, the quantization of the central charge (and not only of the difference $c_{L}-c_{R}$) is also natural from the point of view of the dual conformal field theory. This is because of the Zamolodchikov $c$-theorem \cite{Zamolodchikov1986c}, which states the impossibility of having a family of CFTs with a $SL(2,\mathbb{R})\times SL(2,\mathbb{R})$ invariant vacuum parameterized by a continuous value of the central charge. In turn, consistency of the theory, provided one assumes the AdS/CFT conjecture, demands the dimensionless ratios constructed by the different coupling constants of the theory to take special values for the bulk theory to be well defined.

Furthermore, one could also ask whether there is a way to understand these quantization conditions from the point of view of the microscopic theory. To analyze this, one could think of embedding the
three-dimensional gravity action, including the exotic Chern-Simons term, in a bigger consistent theory, like string theory. Even though a complete description of it has not yet been accomplished (see \cite{Lu2010h} for a recent attempt), one can consider a toy example to see how it would work. For instance, let us play around with the $\mathcal{O}(R^{4})$ M-theory terms, which are those that supplement the eleven-dimensional supergravity action. Among such higher-curvature terms one finds couplings between the 3-form $A=A_{\mu \nu \rho }\ dx^{\mu }\wedge dx^{\nu }\wedge dx^{\rho}$ with the curvature tensor $R_{\ \mu \nu }^{ab}=e_{\alpha }^{a}e_{\beta }^{b}R_{\ \mu\nu }^{\alpha \beta }$.

One such term is of the form $\int_{\Sigma_{11}}A\wedge ${tr}$(R\wedge R)\wedge ${tr}$(R\wedge R)$, together with other terms (the trace is taken over the indices in the tangent bundle $a,b,...$). Then, one can think of a compactification of the form\footnote{G.G. thanks B.S. Acharya for suggesting the possibility of this type of construction.} $\Sigma_{11}=\Sigma _{3}\times M_{4}\times X_{4}$, with $F=dA$ having flux on $M_{4}$, and asking $X_{4}$ to have non-trivial signature (non-vanishing Pontryagin invariant). Integrating by parts the higher-order term written above, one finds a contribution of the form $-\int_{\Sigma _{3}}(\omega_{a}\wedge d\omega ^{a}+\frac{\varepsilon _{abc}}{3}\omega ^{a}\wedge \omega^{b}\wedge \omega ^{c})$ $\int_{M_{4}}F${\ }$\int_{X_{4}}R_{ab}\wedge R^{ab}$, so that the exotic gravitational term appears here, being the effective three-dimensional coupling $(8\pi G\mu )^{-1}\sim $ $\sigma_{(X_{4})}N_{(M_{4})}$, where $\sigma _{(X_{4})}$ is the signature of $X_{4}$ and $N_{(M_{4})}$ is the charge under $F$. This sketches how a (yet to be found) microscopic realization could yield the quantization condition for $c_{R}-c_{L}$.

\subsection{Black hole entropy}

Now, before concluding the discussion on the central charge, let us consider a quick application of the result (\ref{cRcL}). The values of the central charges derived above provide us with a tool to compute the black hole entropy microscopically. This was discussed in \cite{Blagojevic2006,Blagojevic2006b,Blagojevic2006a} for the case of the theory with torsion, and it follows the well-known procedure originally proposed by Strominger in \cite{Strominger1998a}. This amounts to considering the Cardy formula \cite{Cardy1986} of the dual CFT. In a two-dimensional CFT, Cardy's formula gives an asymptotic expression for the growing of density of states. It follows from modular invariance and some general hypothesis about the spectrum of the theory. The formula for the microcanonical entropy, representing the logarithm of the number of degrees of freedom for given values of $\mathcal{M}$ and $\mathcal{J}$, reads
\begin{equation}
S_{\text{CFT}}=2\pi \sqrt{\frac{c_{L}}{12}\left( \mathcal{M}l-\mathcal{J}\right) }+2\pi \sqrt{\frac{c_{R}}{12}\left( \mathcal{M}l+\mathcal{J}\right) } ~,  \label{entropy2}
\end{equation}
where the conserved charges associated to Killing vectors $\partial _{t}$ and $\partial _{\phi }$, namely the mass and the angular momentum, are identified with the Virasoro generators $L_{0}+\overline{L}_{0}$ and $L_{0}-\overline{L}_{0}$ respectively. Resorting to equations (\ref{MJ2}), (\ref{horizontes}) and (\ref{cRcL}), one actually verifies that (\ref{entropy2}) exactly reproduces the black hole entropy (\ref{entropy1}); see \cite{Blagojevic2006,Blagojevic2006b,Blagojevic2006a,Klemm2008,Cacciatori2006}.

\section{A singular limit}

\subsection{Degeneracy in Mielke-Baekler theory}

Now, let us consider the special case $\mu =m=\sqrt{-\Lambda }$. As said before, in this case the equations of motion (\ref{eom2}) and (\ref{2eom}) coincide and the Mielke-Baekler theory develops a kind of degeneracy as the equations of motion only give
\begin{equation}
R^{a}+\mu T^{a}+\frac{\mu ^{2}}{2}\varepsilon _{\ bc}^{a}e^{b}\wedge e^{c}=0 ~.
\label{consistency}
\end{equation}
Certainly, this equation is not sufficiently restrictive unless one specifies additional information, e.g. about the torsion. On the other hand, if $\mu =m$ equations (\ref{eom2}) and (\ref{2eom}) can not be generically written in the form (\ref{Rcte}) and (\ref{Tcte}). In fact, $\mu =m=\sqrt{-\Lambda }$ is a singular point of the theory. This is why in order to analyze this point it is necessary to take the limit carefully proposing a consistent prescription. A particular consistent way this limit can be taken is to actually consider the form (\ref{Rcte}) and (\ref{Tcte}) for the equations of motion, namely
\begin{equation}
R^{a}=\frac{\mu }{2}\frac{\Lambda +m^{2}}{\mu -m}\,\varepsilon_{\,\,\,bc}^{\,a}\,e^{b}\wedge e^{c} ~, \qquad
T^{a}=\frac{1}{2}\frac{m\mu+\Lambda}{m-\mu}\,\varepsilon_{\,\,\,bc}^{\,a}\,e^{b}\wedge e^{c} ~,
\label{cW}
\end{equation}
define $1-m/\mu =\varepsilon $, and then take the limit $\varepsilon $ going to zero in such a way that the equations (\ref{cW}) remain well defined. For this to be consistent one has to consider the limit $1-m/\mu =\varepsilon \rightarrow 0$ together with the limit $\Lambda +m^{2}=\varepsilon
/l^{2}\rightarrow 0$. Then, if the torsion is set to zero, (\ref{consistency}) would require $-1/l^{2}$ to coincide with the constant $\Lambda $ appearing in the Lagrangian, and so one finds that $m+\Lambda /\mu $ identically vanishes. In turn, the limit $1-m/\mu \rightarrow 0$ is consistent with (\ref{cW}) and one eventually obtains
\begin{equation}
R^{a}=\frac{\Lambda }{2}\,\varepsilon _{\,\,\,bc}^{\,a}\,e^{b}\wedge e^{c},\qquad T^{a}=0 ~;
\label{A1}
\end{equation}
that is, the Einstein equations. In this limit one also finds the central charges
\begin{equation}
c_{L}=0 ~, \qquad \qquad c_{R}=\frac{3l}{G} ~,  \label{A2}
\end{equation}
with $l^{-2}=-\Lambda $. Finally, to take the analogy with the model of \cite{Li2008} one step further, one may notice that at this special point all the black hole solutions of the theory fulfill the extremal relation
\begin{equation}
l\mathcal{M}=\mathcal{J} ~. \label{A3}
\end{equation}
On the other hand, it seems clear that we could have also taken the $\mu \rightarrow m$ limit in such a way that it is the torsion the quantity that does not vanish at the critical point, obtaining, instead of (\ref{A1}), the following
\begin{equation}
T^{a}=\frac{1}{2}\beta\,\varepsilon_{\,\,\,bc}^{\,a}\,e^{b}\wedge e^{c}, \label{torbeta}
\end{equation} 
with arbitrary value $\beta$. In this case, the effective cosmological constant would have been given by $-l^{-2}=\Lambda (1+\beta)$, and then we would have ended up having a non-vanishing torsion at the critical point. That is, the point $\mu =m=\sqrt{-\Lambda }$ is a degenerate point of Mielke-Baekler theory and such degeneracy gets realized by the ambiguity in the choice of $\beta$, which is fixed only after a particular prescription for the limit is adopted. The choice $\beta = 0$ gives a theory similar to that pursued in \cite{Li2008}. Besides, it is clear from (\ref{consistency}) that at the degenerate point the theory neither gives information about the curvature nor about the torsion, but about the combination $R^{a}+\mu T^{a}$. Then, the only equation of motion written in the Chern-Simons form turns out to be $F=0$, which is a field equation for $A^{a}=\omega^{a}+\mu \ e^{a}$. Here, it is worth emphasizing that, at $m = \mu = \sqrt{-\Lambda }$, the theory defined by (\ref{consistency}) and that defined by (\ref{cW}) are not equivalent. This is the case even when the systems of equations (\ref{Eom2})-(\ref{Eom3}) and (\ref{EOM1})-(\ref{EOM2}) are equivalent if $m\neq \mu $. In fact, while equations (\ref{cW}) in the limit $m\to \mu \to \sqrt{-\Lambda}$ still define a theory with constant curvature and constant torsion, equation (\ref{consistency}) only gives information about the quantity $R^a +\mu T^a$. It is (\ref{consistency}), and not (\ref{cW}), the model that corresponds to a single Chern-Simons field theory. 

The singular point, as we will shortly analyze in the next subsection within the canonical formalism, gives a particular combination of the coupling constants for which some of the would be degrees of freedom simply decouple (the situation here is a bit more cumbersome since these theories have no local degrees of freedom on their own). If the microscopic Lagrangian of the theory is fine tuned to those values that would lead to the critical point, one should simply make a field redefinition from scratch and the theory becomes a Chern-Simons theory for a single $SL(2,\mathbb{R})$, whose geometrical meaning is unclear. However, whatever approach to this problem is chosen, it seems more natural to embed the Mielke-Baekler Lagrangian into a bigger picture, so that the singular point is eventually approached to from the generic non-degenerate situation. As such, it is natural to give a prescription for the singular point that smoothly interpolates with the generic case, where both the curvature and the torsion are constant. Still, there is some freedom within this prescription, which is reflected in the parameter $\beta$ in (\ref{torbeta}). The choice $\beta = 0$ is special in that it makes the theory closely reminiscent to chiral gravity \cite{Li2008}.

To understand the indefinition in the parameter $\beta $, it is worth studying the map between different geometries and how it behaves at the degenerate point: In Mielke-Baekler theory there is a natural way to establish a map between geometries which are solutions of the theory (\ref{SMB}) for different values of the coupling constants \cite{Giacomini2007}. That is, one can perform a linear transformation of the fields like
\begin{equation}
\omega ^{a}\rightarrow \omega ^{a}+\beta \ e^{a} ~,\qquad e^{a}\rightarrow e^{a} ~.
\label{beta}
\end{equation}
and find that this transformation induces a transformation of the four coupling constants that appear in the action. To give an example of how it works, it is sufficient to consider the Lagrangian of the theory in the particular case in which its coupling constants satisfy the relation $\mu m=-\Lambda$. In this case, a transformation like (\ref{beta}) generates the following
transformation of the coupling constants
\begin{eqnarray}
G &\rightarrow &\widetilde{G}=\frac{G\mu }{\mu +\beta} ~, \\ [0.7em]
\mu  &\rightarrow &\widetilde{\mu }=\mu +\beta ~, \\ [0.9em]
m &\rightarrow &\widetilde{m}=\frac{m\mu +2\mu \beta +\beta ^{2}}{\mu +\beta} ~, \\ [0.7em]
\Lambda  &\rightarrow &\widetilde{\Lambda }=\frac{\Lambda \mu -3m\mu \beta
-3\mu \beta ^{2}-\beta ^{3}}{\mu +\beta} ~.
\end{eqnarray}

The case we started with already satisfies the special condition $\mu m=-\Lambda $, and provided it also satisfies $\mu =m$ one finds that the transformed coupling constants obey $\widetilde{\mu }\widetilde{m}=-\widetilde{\Lambda }$ and $\widetilde{\mu }=\widetilde{m}$ as well. That is,
the special condition $m^{2}=\mu ^{2}=-\Lambda $ appears to be a fixed point of the $\beta $-transformation (\ref{beta}); in fact, after the transformation one finds $\widetilde{\mu }^{2}=\widetilde{m}^{2}=-\widetilde{\Lambda}=(\beta +\mu )^{2}$. And we see that this transformation generates (constant) torsion $T^{a}\sim \beta \ \varepsilon _{\ bc}^{a}e^{b}\wedge e^{c}$ from a configuration with vanishing torsion. The combination that remains invariant is, precisely, $R^{a}+\mu T^{a}+\frac{\mu ^{2}}{2}\varepsilon _{\ bc}^{a}e^{b}\wedge e^{c} \to R^{a}+\tilde{\mu} T^{a}+\frac{\tilde{\mu} ^{2}}{2}\varepsilon _{\ bc}^{a}e^{b}\wedge e^{c} $, as in (\ref{A1}). This explains the degenerate point appearing as a fixed point of (\ref{beta}).

\subsection{Analogy with chiral gravity}

Equations (\ref{A1}), (\ref{A2}) and (\ref{A3}) are actually evocative of what happens in chiral gravity. The point $\mu =m=\sqrt{-\Lambda}$ corresponds to the point of the space of parameters where the Chern-Simons coupling $\hat{q}$ vanishes. In turn, the theory consists of a single Chern-Simons action (see \cite{Witten2010} for a brief comment about the relation between the singular point $\hat{q}=0$ and the chiral point of \cite{Li2008}; cf. \cite{Cacciatori2006}). When $\hat{q}=0$ the left-handed degrees of freedom are left unspecified; however, we have just argued that one could consistently demand the torsionless condition $T^{a}=0$ when approaching the singularity.

We have just identified a special (singular) point of Mielke-Baekler theory at which the theory behaves pretty much like chiral gravity of \cite{Li2008}. That is, it gives a model of three-dimensional gravity that fulfill the following properties:
\begin{itemize}
\item[\textit{a)}] Once suitable asymptotically AdS$_{3}$ boundary conditions are imposed, the asymptotic isometry group turns out to be generated by one (right-handed) Virasoro algebra with central charge $c_{R}=3l/G$, while the central charge of the left-handed part $c_{L}$ vanishes.
\item[\textit{b)}] The theory can be written as a single $SL(2,\mathbb{R})$ Chern-Simons term, as the other copy of the bulk action decouples in the limit, being proportional to $c_{L}$.
\item[\textit{c)}] The BTZ black holes have mass $\mathcal{M}$ and angular momentum $\mathcal{J}$ that obey the relation $l\mathcal{M}=\mathcal{J}$, no matter the values that the parameters $M$ and $J$ of the solution take.
\item[\textit{d)}] The theory has no local degrees of freedom, as it corresponds to a special case of the Mielke-Baekler theory.
\item[\textit{e)}] If the limit $\mu \rightarrow m\rightarrow \sqrt{-\Lambda }$ is taken in such a way that equations (\ref{cW}) are obeyed, all the solutions of the theory at the special point have vanishing torsion, for $\beta = 0$, and are Einstein manifolds, {\it i.e.}, spaces locally AdS$_{3}$.
\end{itemize}

Nevertheless, besides the resemblance between the chiral model obtained from the degenerate case of Mielke-Baekler theory and the chiral gravity of \cite{Li2008}, it is worth emphasizing that both constructions are radically different, for instance in what regards to the property \textit{e)} listed above \cite{Compere2010}.

In the next section, we will analyze the canonical structure of the theory and how it changes at the degenerate point.

\subsection{Canonical analysis}

In the previous section we discussed a degenerate point of the Mielke-Baekler theory of gravity in AdS$_{3}$ space and we proposed a prescription to approach this point in the space of parameters. Now, let us briefly discuss the canonical structure of the theory. Our discussion will follow the approach and notation of references \cite{Blagojevic2004,Banerjee2010a}, but paying special attention to the analysis of the constrained system in order not to miss the difference between the critical and the non-critical cases.

The Hamiltonian analysis of the theory starts by slicing the three-dimensional spacetime manifold, separating the temporal components from the spatial ones, and defining a configuration space. The coordinates of this configuration space (henceforth denoted by $q$) are the components $e_{a}^{\mu }$ and $\omega _{a}^{\mu }$. Explicitly we can write the canonical momenta associated to these as follows
\begin{equation}
\pi_{a}^{0}=0 ~, \quad \Pi _{a}^{0}=0 ~, \quad \pi _{a}^{i}=\mu \varepsilon ^{ij}\left( \omega_{aj}+m\ e_{aj}\right) ~, \quad \Pi _{a}^{i}=\varepsilon ^{ij}\left( \omega_{aj}+\mu \ e_{aj}\right) ~,
\end{equation}
which correspond to $e_{a}^{0}$, $\omega _{a}^{0}$, $e_{a}^{i}$ and $\omega_{a}^{i}$, respectively, where the notation is such that $i,j=1,2$ refer to the spatial part of the spacetime indices. The canonical momenta are indeed defined with respect to the action (\ref{SMB}) times $16 \pi G \mu$. These relations define the primary constraints of the theory; namely 
\begin{equation}
\phi_{a}^{0} \equiv \pi_{a}^{0} ~, \quad
\Phi_{a}^{0} \equiv \Pi _{a}^{0} ~, \quad
\phi_{a}^{i} \equiv \pi _{a}^{i}-\mu \varepsilon ^{ij}\left( \omega _{aj}+m e_{aj}\right) ~, \quad
\Phi _{a}^{i}\equiv \Pi _{a}^{i}-\varepsilon ^{ij}\left( \omega_{aj}+\mu e_{aj}\right) ~.
\end{equation}
Then, the primary Hamiltonian density is
\begin{equation}
\cH_{T}=e_{0}^{a}\,\cH_{a}+\o_{0}^{a}\,\cK_{a}+\dot{e}_{0}^{a}\,\p_{a}^{0}+\dot{\o}_{0}^{a}\,\P_{a}^{0}+\dot{e}_{i}^{a}\,\p_{a}^{i}+\dot{\o}_{i}^{a}\,\P_{a}^{i} ~,
\end{equation}
where the dot stands for time derivatives and, following the notation used in \cite{Blagojevic2004}, 
\begin{eqnarray}
\cH^{a}& =-\m\left( m\, T_{ij}^{a}+R_{ij}^{a}-\Lambda\,\varepsilon_{\phantom{a}bc}^{a}e_{i}^{b}e_{j}^{c}\right) \varepsilon^{ij} ~, \\ [0.7em]
\cK^{a}& =-\left( \m T_{ij}^{a}+R_{ij}^{a}+m\,\mu\,\varepsilon_{\phantom{a}bc}^{a}e_{i}^{b}e_{j}^{c}\right) \varepsilon^{ij} ~.
\end{eqnarray}
The dynamics of the theory is generated by $\cH_{T}$, while the time derivatives of the coordinates that accompany the constraints play the r\^{o}le of Lagrange multipliers that fix them to zero. The structure of the Hamiltonian is, in general, given by $\cH_{T}=\tilde{\cH}+\dot{q}^{I}\,\p_{I}$.  That is, the actual Hamiltonian is given by the sum of the canonical Hamiltonian and the contributions coming from the constraints. The Poisson structure arises from imposing canonical
constraints on coordinates and momenta through the Lie bracket $\{,\}$. The constraints $\p_I=0$ reduce the original phase space to the physical one, and consistency of the theory demands the constraints to be preserved through the dynamical evolution of the system in the reduced phase space. This requires $\dot{\p}_{J}$ to weakly vanish,
\begin{equation}
\dot{\p}_{J}=\{\cH_{T},\p_{J}\}=\{\tilde{\cH},\p_{J}\}+\dot{q}^{I}\{\p_{I},\p_{J}\} \approx 0 ~.
\end{equation}
In our case, we have
\begin{eqnarray}
&\dot{\p}_{a}^{0} = -\cH_{a} ~, \qquad \dot{\P }_{a}^{0}=-\cK_{a} ~, \nonumber \\ [0.9em]
& \dot{\p}_{a}^{i} =2\m m \epsilon ^{ji} \left( \pa_j e_{a0}-\epsilon_{ab}^{\phantom{ab}c}\left(\o^b_j-\displaystyle\frac{\L}{m}e^b_j\right)e_{c0} \right)+ 2\m \epsilon ^{ji} \left( \pa_j \o_{a0}-\epsilon_{ab}^{\phantom{ab}c}\left(\o^b_j+m e^b_j\right)\o_{c0} \right) \nonumber\\ [0.9em]
&+2\m\epsilon ^{ji}\left( m\dot{e}_{aj}+\dot{\o}_{aj}\right) , \nonumber \\ [0.9em]
& \dot{\P }_{a}^{i} =2\m \epsilon ^{ji} \left( \pa_j e_{a0}-\epsilon_{ab}^{\phantom{ab}c}\left(\o^b_j+m e^b_j\right)e_{c0} \right)+ 2 \epsilon ^{ji} \left( \pa_j \o_{a0}-\epsilon_{ab}^{\phantom{ab}c}\left(\o^b_j+\m e^b_j\right)\o_{c0} \right) \nonumber\\ [0.9em]
&+2\epsilon ^{ji}\left( \m\dot{e}_{aj}+\dot{\o}_{aj}\right) ~, \nonumber
\end{eqnarray}
The first line above expresses the fact that that $\cH_{a}$ and $\cK_{a}$ are secondary constraints, while the second and third lines give equations that allows to find the values of $\dot{e}_{aj}$ and $\dot{\o }_{aj}$ that set these expressions to zero. Solving these equations is always possible except when the determinant of the system is zero, what precisely occurs when $m=\m$. Leaving the critical case aside for a moment, one can continue the analysis and verify that the secondary constraints are actually consistent: the non-trivial Poisson brackets for $m\neq \mu$ are 
\begin{equation}
\poi{\p^i_a}{\p^j_b} =-2m\mu\,\varepsilon^{ij}\,\delta_{ab} ~, \quad
\poi{\p^i_a}{\P^j_b} =-2\mu\,\varepsilon^{ij}\,\delta_{ab} ~, \quad
\poi{\P^i_a}{\P^j_b} =-2\,\varepsilon ^{ij}\,\delta_{ab} ~,
\label{eq:firstcommutator}
\end{equation}
with
\begin{eqnarray}
& & \poi{\p^i_a}{\bar{\cH}_b} = \varepsilon _{ab}^{\phantom{ab}c}\left( \displaystyle\frac{\Lambda +m\mu }{m-\mu}\,\p_{c}^{i}+\mu \displaystyle\frac{\Lambda +m^{2}}{\mu -m}\,\P_{c}^{i}\right) ~, \\ [0.9em]
& & \poi{\p^i_a}{\bar{\cK}_b} = \poi{\P^i_a}{\bar{\cH}_b} =-\varepsilon _{ab}^{\phantom{ab}c}\p_{c}^{i} ~, \quad \poi{\P^i_a}{\bar{\cK}_b} = -\varepsilon _{ab}^{\phantom{ab}c}\P_{c}^{i} ~,
\end{eqnarray}
and
\begin{eqnarray}
& & \poi{\bar{\cH}_a}{\bar{\cH}_b} = \varepsilon_{ab}^{\phantom{ab}c}\left( \displaystyle\frac{\Lambda +m\mu }{m-\mu}\,\bar{\cH}_{c}+\mu \displaystyle\frac{\Lambda +m^{2}}{\mu -m}\,\bar{\cK}_{c}\right) ~, \label{eq:lastcommutator} \\ [0.9em]
& & \poi{\bar{\cH}_a}{\bar{\cK}_b} =-\varepsilon _{ab}^{\phantom{ab}c}\bar{\cH}_{c} ~, \quad \poi{\bar{\cK}_a}{\bar{\cK}_b} =-\varepsilon _{ab}^{\phantom{ab}c}\bar{\cK}_{c} ~.
\end{eqnarray}
There is of course a $\delta^{2}(\vec x - \vec y)$ implicit in all these formulas. After substituting the expression for the multipliers back into the total Hamiltonian, one can integrate by parts to rearrange the factors that accompany the canonical variables, that instead of $\cH$ and $\cK$ are now,
\begin{eqnarray}
&\bar{\cH}_a =\cH_a-\left(\pa_i\phi^i_a-\epsilon_{ab}^{\phantom{ab}c}\,\o^b_i\,\phi^i_c\right)- \epsilon_{ab}^{\phantom{ab}c}\,e^b_i\left( \displaystyle\frac{\L+m\m}{m-\m}\;\phi^i_c+\m \frac{\L+m^2}{\m-m}\Phi^i_c\right) ~, \nonumber \\[0.9em]
&\bar{\cK}_a =\cK_a-\left(\pa_i\Phi^i_a-\epsilon_{ab}^{\phantom{ab}c}\,\o^b_i\,\Phi^i_c\right)+\epsilon_{ab}^{\phantom{ab}c}\,e^b_i\,\Phi^i_c ~. \nonumber
\end{eqnarray}

In contrast, at the critical point the theory exhibits a dynamical pathology. The reason is that, when $m =\mu $, a new symmetry appears, and this must be properly taken into account when analyzing the constraints. What happens when going from the generic case to the critical case $m=\mu $ is that two of the momenta become proportional to each other, namely $\pi_{a}^{i}=\m\,\Pi_{a}^{i}$, and consequently the respective constraints happen to carry the same information. This is basically because at such point of the space of parameters the coordinates $e^{a}$ and $\omega^{a}$ play symmetric r\^{o}les in the action. As mentioned before, at the singular point one of the Chern-Simons actions drops out and one is left with a single action describing the dynamics of the field $A^a =\omega ^{a}+\mu e^{a}$. In order to take this symmetry (between the r\^{o}le played by $e^{a}$ and $\omega ^{a}$) into account, one can replace the constraint $\p_{a}^{i}$ by the new one $\psi_{a}^{i}\equiv \p_{a}^{i}-\m\P _{a}^{i}$, in such a way that the constraints turn out to be given by
\begin{eqnarray}
& & \dot{\p}_{a}^{0} =-\cJ_{a} ~, \quad \dot{\P }_{a}^{0}=-\cJ_{a} ~, \quad \dot{\psi}_{a}^{i} = 0~, \label{eq:multipliers3} \\ [0.9em]
& & \dot{\P}_{a}^{i} = 2\epsilon^{ij}\left(\pa_jA_{a0}-\epsilon_{ab}^{\phantom{ab}c}A^b_j A_{c0}\right)+2\epsilon ^{ji}\left( \m\dot{e}_{aj}+\dot{\o }_{aj}\right) ~,
\end{eqnarray}
where the last equation can always be solved. $\cJ_{a}=\cH_{a}/\m=\cK_{a}=-\epsilon^{ij}F^a_{ij}$ at the critical point, where $\m=m=\sqrt{-\Lambda}$. The non-zero Poisson brackets in the critical case are
\begin{equation}
\poi{\P^i_a}{\P^j_b} =-2\varepsilon ^{ij}\delta _{ab} ~, \quad
\poi{\P^i_a}{\bar{\cJ}_b} =-\varepsilon _{ab}^{\phantom{ab}c}\P_{c}^{i} ~, \quad
\poi{\bar{\cJ}_a}{\bar{\cJ}_b} =-\varepsilon _{ab}^{\phantom{ab}c}\bar{\cJ}_{c} ~,
\end{equation}
where
\begin{equation}
\bar{\cJ}_a=\cJ_a-\left(\pa_i \Phi^i_a-\epsilon_{ab}^{\phantom{ab}c}A^b_i\Phi^i_c\right).
\end{equation}
The difference between the critical point $m=\mu$ and the generic case can be summarized easily by counting the amount of constraints of first class (FC) and of second class (SC) that appear in each case. Namely\vskip5mm

\begin{minipage}[t]{.45\linewidth}
\begin{center}
\begin{tabular}{|c|c|c|}
\hline
 & Primary & Secondary \\
\hline
FC & $\p^0_a$,  $\P^0_a$ & $\bar{\cH}_a$, $\bar{\cK}_a$ \\
\hline
SC & $\p^i_a$, $\P^i_a$ & ----- \\
\hline
\end{tabular}
\end{center}
\end{minipage}
\begin{minipage}[t]{.45\linewidth}
\begin{center}
\begin{tabular}{|c|c|c|}
\hline
 & Primary & Secondary \\
\hline
FC & $\p^0_a$,  $\P^0_a$, $\psi^i_a$ & $\bar{\cJ}_a$ \\
\hline
SC & $\P^i_a$ & ----- \\
\hline
\end{tabular}
\end{center}
\end{minipage}\vskip5mm

We see that at the critical point one primary constraint of second class is promoted to the first class\footnote{Notice that the number of degrees of freedom in each case remains to be zero.}, and the two secondary constraints of the first class, $\cH$ and $\cK$, collapse to one, denoted by $\cJ$. This indicates that a new symmetry appears at the critical point; namely 
\begin{equation}
{\delta_{\xi }}e_{i}^{a}=\xi_{i}^{a} ~, \qquad {\delta_{\xi }}\o _{i}^{a}=-\m\,\xi_{i}^{a} ~,
\end{equation}
which certainly leaves $A_{i}^{a}$ invariant. 

It is worth noticing that the prescription for going from the general case to the critical point defined through the limiting procedure $1-m/\m =\varepsilon $, $\L -m^{2}=\varepsilon /l^{2}$, and $m+\L /\m=0$ can be applied to the set of commutators (\ref{eq:firstcommutator})-(\ref{eq:lastcommutator}), along with the change $\p_{a}^{i}\rightarrow \psi_{a}^{i}=\p_{a}^{i}-\m\P _{a}^{i}$ and $\cJ_{a}=\cH_{a}/\m=\cK_{a}$ to obtain the set of commutation relations of the critical case. This can also be done for arbitrary value of the parameter $\beta$, introduced earlier, without distinction.

\section{Conclusions}

We have considered the Mielke-Baekler theory of gravity in asymptotically AdS%
$_{3}$ spacetime with torsion. We have reviewed the computation of central charges of the asymptotic
algebra, which turn out to be the central charges of the dual CFT$_{2}$. The result we obtained agrees with the central charges obtained in the literature by employing different methods \cite{Blagojevic2003a,Blagojevic2003,Blagojevic2003b,Klemm2008,Cacciatori2006}. It was observed that a special point of the space of parameters exists, at which one of the central charges vanishes. This point was compared with the chiral point of topologically massive gravity, and the analogies between both models were pointed out. This point is a singular point for the Mielke-Baekler theory, where the theory exhibits degeneracy. We analyzed this at the level of Chern-Simons theory and in the canonical approach. In the Chern-Simons formulation this critical point appears as the point of
the space of parameters at which one of the two $SL(2,\mathbb{R})$ actions drops out. This point was recently mentioned in \cite{Witten2010} within the context of the analytically extended theory, where the connection with the chiral gravity of \cite{Li2008} was already mentioned. It was one of our motivations to make this connection with chiral gravity more explicit. 

One of the aspects one observes here is that several features of the dual conformal field theory do not seem to depend on the precise prescription adopted to reach the singular point of the Mielke-Baekler theory. This raises the question as to whether the relevant physical information is independent of the way one approaches $m=\mu =\sqrt{-\Lambda}$. Despite quantities in the geometric realization do actually depend on how the limit is taken, this possibly reflecting the fact that the theory becomes in essence non-geometrical, it seems plausible that all these geometries are different realizations of the same theory, and, likely, the way of making sense out of Mielke-Baekler theory at the point it exhibits degeneracy is, in fact, resorting to the dual description in terms of a chiral CFT.

\section*{Acknowledgements}


This note is based on talks that some of the authors delivered in 2009 and 2010 at different institutions. They thank Universit\'{e} Libre de Bruxelles, Brussels; CERN, Geneva; Brandeis University, Boston; Universit\'{e} Pierre et Marie Curie, Paris; Abdus Salam International Centre for Theoretical Physics, Trieste; Universitat de Barcelona, Catalunya; Universidad de Santiago de Compostela, Galicia; Centro de Estudios Cient\'{\i}ficos, Valdivia; Erwin Schr\"{o}dinger Institute, Vienna; Institute of Physics and Mathematics of the Universe, Tokyo, for the hospitality.
Discussions with\ B. Acharya, G. Barnich, S. Detournay, J. Evslin, D. Grumiller, N. Johansson, M. Kleban, M. Leoni, M. Leston, O. Mi\v{s}kovi\'{c}, R. Olea, J. Oliva, M. Porrati, R. Troncoso, and J. Zanelli are acknowledged.

This work was supported in part by ANPCyT, CONICET and UBA (Argentina), by MICINN and FEDER (grant FPA2008-01838), by Xunta de Galicia (Conseller\'{\i}a de Educaci\'on and grant PGIDIT10PXIB206075PR), by the Spanish Consolider-Ingenio 2010 Programme CPAN (CSD2007-00042), and by a bilateral agreement MINCyT Argentina (ES/08/02) -- MICINN Spain (FPA2008-05138-E).
The Centro de Estudios Cient\'\i ficos (CECS) is funded by the Chilean Government through the Millennium Science Initiative and the Centers of Excellence Base Financing Program of Conicyt, and by the Conicyt grant ``Southern Theoretical Physics Laboratory'' ACT-91. CECS is also supported by a group of private companies which at present includes Antofagasta Minerals, Arauco, Empresas CMPC, Indura, Naviera Ultragas and Telef\'onica del Sur.
%


\begin{thebibliography}{10}

\bibitem{Li2008}
W.~Li, W.~Song, and A.~Strominger, {\it {Chiral Gravity in Three Dimensions}},
  {\em JHEP} {\bf 04} (2008) 082,
  [\href{http://xxx.lanl.gov/abs/0801.4566}{{\tt arXiv:0801.4566}}].

\bibitem{Strominger2008}
A.~Strominger, {\it {A Simple Proof of the Chiral Gravity Conjecture}},
  \href{http://xxx.lanl.gov/abs/0808.0506}{{\tt arXiv:0808.0506}}.

\bibitem{Carlip2008}
S.~Carlip, S.~Deser, A.~Waldron, and D.~K. Wise, {\it {Topologically Massive
  AdS Gravity}},  {\em Phys. Lett.} {\bf B666} (2008) 272--276,
  [\href{http://xxx.lanl.gov/abs/0807.0486}{{\tt arXiv:0807.0486}}].

\bibitem{Carlip2009}
S.~Carlip, S.~Deser, A.~Waldron, and D.~K. Wise, {\it {Cosmological
  Topologically Massive Gravitons and Photons}},  {\em Class. Quant. Grav.}
  {\bf 26} (2009) 075008, [\href{http://xxx.lanl.gov/abs/0803.3998}{{\tt
  arXiv:0803.3998}}].

\bibitem{Li2008a}
W.~Li, W.~Song, and A.~Strominger, {\it {Comment on 'Cosmological Topological
  Massive Gravitons and Photons'}},
  \href{http://xxx.lanl.gov/abs/0805.3101}{{\tt arXiv:0805.3101}}.

\bibitem{Giribet2008b}
G.~Giribet, M.~Kleban, and M.~Porrati, {\it {Topologically Massive Gravity at
  the Chiral Point is Not Chiral}},  {\em JHEP} {\bf 10} (2008) 045,
  [\href{http://xxx.lanl.gov/abs/0807.4703}{{\tt arXiv:0807.4703}}].

\bibitem{Carlip2008c}
S.~Carlip, {\it {The Constraint Algebra of Topologically Massive AdS Gravity}},
   {\em JHEP} {\bf 10} (2008) 078,
  [\href{http://xxx.lanl.gov/abs/0807.4152}{{\tt arXiv:0807.4152}}].

\bibitem{Carlip2009c}
S.~Carlip, {\it {Chiral Topologically Massive Gravity and Extremal B-F
  Scalars}},  {\em JHEP} {\bf 09} (2009) 083,
  [\href{http://xxx.lanl.gov/abs/0906.2384}{{\tt arXiv:0906.2384}}].

\bibitem{Henneaux2009}
M.~Henneaux, C.~Martinez, and R.~Troncoso, {\it {Asymptotically anti-de Sitter
  spacetimes in topologically massive gravity}},  {\em Phys. Rev.} {\bf D79}
  (2009) 081502, [\href{http://xxx.lanl.gov/abs/0901.2874}{{\tt
  arXiv:0901.2874}}].

\bibitem{Grumiller2008}
D.~Grumiller and N.~Johansson, {\it {Instability in cosmological topologically
  massive gravity at the chiral point}},  {\em JHEP} {\bf 07} (2008) 134,
  [\href{http://xxx.lanl.gov/abs/0805.2610}{{\tt arXiv:0805.2610}}].

\bibitem{Grumiller2009}
D.~Grumiller and N.~Johansson, {\it {Consistent boundary conditions for
  cosmological topologically massive gravity at the chiral point}},  {\em Int.
  J. Mod. Phys.} {\bf D17} (2009) 2367--2372,
  [\href{http://xxx.lanl.gov/abs/0808.2575}{{\tt arXiv:0808.2575}}].

\bibitem{Ertl2009}
S.~Ertl, D.~Grumiller, and N.~Johansson, {\it {Erratum to `Instability in
  cosmological topologically massive gravity at the chiral point',
  arXiv:0805.2610}},  \href{http://xxx.lanl.gov/abs/0910.1706}{{\tt
  arXiv:0910.1706}}.

\bibitem{Maloney2010}
A.~Maloney, W.~Song, and A.~Strominger, {\it {Chiral Gravity, Log Gravity and
  Extremal CFT}},  {\em Phys. Rev.} {\bf D81} (2010) 064007,
  [\href{http://xxx.lanl.gov/abs/0903.4573}{{\tt arXiv:0903.4573}}].

\bibitem{Henneaux2010}
M.~Henneaux, C.~Martinez, and R.~Troncoso, {\it {More on Asymptotically Anti-de
  Sitter Spaces in Topologically Massive Gravity}},  {\em Phys. Rev.} {\bf D82}
  (2010) 064038, [\href{http://xxx.lanl.gov/abs/1006.0273}{{\tt
  arXiv:1006.0273}}].

\bibitem{Gaberdiel2010}
M.~R. Gaberdiel, D.~Grumiller, and D.~Vassilevich, {\it {Graviton 1-loop
  partition function for 3-dimensional massive gravity}},  {\em JHEP} {\bf 11}
  (2010) 094, [\href{http://xxx.lanl.gov/abs/1007.5189}{{\tt
  arXiv:1007.5189}}].

\bibitem{Cunliff2010}
C.~Cunliff, {\it {Topologically Massive Gravity from the Outside In}},
  \href{http://xxx.lanl.gov/abs/1012.2180}{{\tt arXiv:1012.2180}}.

\bibitem{Compere2010}
G.~Compere, S.~de~Buyl, and S.~Detournay, {\it {Non-Einstein geometries in
  Chiral Gravity}},  {\em JHEP} {\bf 10} (2010) 042,
  [\href{http://xxx.lanl.gov/abs/1006.3099}{{\tt arXiv:1006.3099}}].

\bibitem{Mardones1991}
A.~Mardones and J.~Zanelli, {\it {Lovelock-Cartan theory of gravity}},  {\em
  Class. Quant. Grav.} {\bf 8} (1991) 1545--1558. \bibitem{Mielke1991}
E.~W. Mielke and P.~Baekler, {\it {Topological gauge model of gravity with
  torsion}},  {\em Phys. Lett.} {\bf A156} (1991) 399--403.

\bibitem{Maldacena1998}
J.~M. Maldacena, {\it {The large N limit of superconformal field theories and
  supergravity}},  {\em Adv. Theor. Math. Phys.} {\bf 2} (1998) 231--252,
  [\href{http://xxx.lanl.gov/abs/hep-th/9711200}{{\tt hep-th/9711200}}].

\bibitem{Witten1998}
E.~Witten, {\it {Anti-de Sitter space and holography}},  {\em Adv. Theor. Math.
  Phys.} {\bf 2} (1998) 253--291,
  [\href{http://xxx.lanl.gov/abs/hep-th/9802150}{{\tt hep-th/9802150}}].

\bibitem{Gubser1998}
S.~S. Gubser, I.~R. Klebanov, and A.~M. Polyakov, {\it {Gauge theory
  correlators from non-critical string theory}},  {\em Phys. Lett.} {\bf B428}
  (1998) 105--114, [\href{http://xxx.lanl.gov/abs/hep-th/9802109}{{\tt
  hep-th/9802109}}].

\bibitem{Coussaert1995}
O.~Coussaert, M.~Henneaux, and P.~van Driel, {\it {The Asymptotic dynamics of
  three-dimensional Einstein gravity with a negative cosmological constant}},
  {\em Class. Quant. Grav.} {\bf 12} (1995) 2961--2966,
  [\href{http://xxx.lanl.gov/abs/gr-qc/9506019}{{\tt gr-qc/9506019}}].

\bibitem{Henneaux2000}
M.~Henneaux, L.~Maoz, and A.~Schwimmer, {\it {Asymptotic dynamics and
  asymptotic symmetries of three- dimensional extended AdS supergravity}},
  {\em Annals Phys.} {\bf 282} (2000) 31--66,
  [\href{http://xxx.lanl.gov/abs/hep-th/9910013}{{\tt hep-th/9910013}}].

\bibitem{Blagojevic2003a}
M.~Blagojevic and M.~Vasilic, {\it {Asymptotic symmetries in 3d gravity with
  torsion}},  {\em Phys. Rev.} {\bf D67} (2003) 084032,
  [\href{http://xxx.lanl.gov/abs/gr-qc/0301051}{{\tt gr-qc/0301051}}].

\bibitem{Blagojevic2003}
M.~Blagojevic and M.~Vasilic, {\it {3D gravity with torsion as a Chern-Simons
  gauge theory}},  {\em Phys. Rev.} {\bf D68} (2003) 104023,
  [\href{http://xxx.lanl.gov/abs/gr-qc/0307078}{{\tt gr-qc/0307078}}].

\bibitem{Blagojevic2003b}
M.~Blagojevic and M.~Vasilic, {\it {Asymptotic dynamics in 3D gravity with
  torsion}},  {\em Phys. Rev.} {\bf D68} (2003) 124007,
  [\href{http://xxx.lanl.gov/abs/gr-qc/0306070}{{\tt gr-qc/0306070}}].

\bibitem{Klemm2008}
D.~Klemm and G.~Tagliabue, {\it {The CFT dual of AdS gravity with torsion}},
  {\em Class. Quant. Grav.} {\bf 25} (2008) 035011,
  [\href{http://xxx.lanl.gov/abs/0705.3320}{{\tt arXiv:0705.3320}}].

\bibitem{Cacciatori2006}
S.~L. Cacciatori, M.~M. Caldarelli, A.~Giacomini, D.~Klemm, and D.~S. Mansi,
  {\it {Chern-Simons formulation of three-dimensional gravity with torsion and
  nonmetricity}},  {\em J. Geom. Phys.} {\bf 56} (2006) 2523--2543,
  [\href{http://xxx.lanl.gov/abs/hep-th/0507200}{{\tt hep-th/0507200}}].

\bibitem{Witten2010}
E.~Witten, {\it {Analytic Continuation Of Chern-Simons Theory}},
  \href{http://xxx.lanl.gov/abs/1001.2933}{{\tt arXiv:1001.2933}}.

\bibitem{Deser1982}
S.~Deser, R.~Jackiw, and S.~Templeton, {\it {Topologically massive gauge
  theories}},  {\em Ann. Phys.} {\bf 140} (1982) 372--411.

\bibitem{Deser1982a}
S.~Deser, R.~Jackiw, and S.~Templeton, {\it {Three-Dimensional Massive Gauge
  Theories}},  {\em Phys. Rev. Lett.} {\bf 48} (1982) 975--978.

\bibitem{Grumiller2008a}
D.~Grumiller, R.~Jackiw, and N.~Johansson, {\it {Canonical analysis of
  cosmological topologically massive gravity at the chiral point}}, in
  ``Fundamental interactions: a memorial volume for Wolfgang Kummer''
  (2009) 363, [\href{http://xxx.lanl.gov/abs/0806.4185}{{\tt arXiv:0806.4185}}].

\bibitem{Miskovic2009a}
O.~Miskovic and R.~Olea, {\it {Background-independent charges in Topologically
  Massive Gravity}},  {\em JHEP} {\bf 12} (2009) 046,
  [\href{http://xxx.lanl.gov/abs/0909.2275}{{\tt arXiv:0909.2275}}].

\bibitem{Brown1986b}
J.~D. Brown and M.~Henneaux, {\it {Central Charges in the Canonical Realization
  of Asymptotic Symmetries: An Example from Three-Dimensional Gravity}},  {\em
  Commun. Math. Phys.} {\bf 104} (1986) 207--226.

\bibitem{Banados1992}
M.~Banados, C.~Teitelboim, and J.~Zanelli, {\it {The Black hole in
  three-dimensional space-time}},  {\em Phys. Rev. Lett.} {\bf 69} (1992)
  1849--1851, [\href{http://xxx.lanl.gov/abs/hep-th/9204099}{{\tt
  hep-th/9204099}}].

\bibitem{Banados1993}
M.~Banados, M.~Henneaux, C.~Teitelboim, and J.~Zanelli, {\it {Geometry of the
  (2+1) black hole}},  {\em Phys. Rev.} {\bf D48} (1993) 1506--1525,
  [\href{http://xxx.lanl.gov/abs/gr-qc/9302012}{{\tt gr-qc/9302012}}].

\bibitem{Ayon-Beato2006}
E.~Ayon-Beato and M.~Hassaine, {\it {Exploring AdS waves via nonminimal
  coupling}},  {\em Phys. Rev.} {\bf D73} (2006) 104001,
  [\href{http://xxx.lanl.gov/abs/hep-th/0512074}{{\tt hep-th/0512074}}].

\bibitem{Ayon-Beato2009}
E.~Ayon-Beato, G.~Giribet, and M.~Hassaine, {\it {Bending AdS Waves with New
  Massive Gravity}},  {\em JHEP} {\bf 05} (2009) 029,
  [\href{http://xxx.lanl.gov/abs/0904.0668}{{\tt arXiv:0904.0668}}].

\bibitem{Grumiller2010d}
D.~Grumiller and I.~Sachs, {\it {AdS$_3$/LCFT$_2$ -- Correlators in
  Cosmological Topologically Massive Gravity}},  {\em JHEP} {\bf 03} (2010)
  012, [\href{http://xxx.lanl.gov/abs/0910.5241}{{\tt arXiv:0910.5241}}].

\bibitem{Grumiller2011}
D.~Grumiller, N.~Johansson, and T.~Zojer, {\it {Short-cut to new anomalies in
  gravity duals to logarithmic conformal field theories}},  {\em JHEP} {\bf 01}
  (2011) 090, [\href{http://xxx.lanl.gov/abs/1010.4449}{{\tt
  arXiv:1010.4449}}].

\bibitem{Garbarz2009}
A.~Garbarz, G.~Giribet, and Y.~Vasquez, {\it {Asymptotically AdS$_3$ Solutions
  to Topologically Massive Gravity at Special Values of the Coupling
  Constants}},  {\em Phys. Rev.} {\bf D79} (2009) 044036,
  [\href{http://xxx.lanl.gov/abs/0811.4464}{{\tt arXiv:0811.4464}}].

\bibitem{Nieh1982}
H.~T. Nieh and M.~L. Yan, {\it {An identity in Riemann--Cartan geometry}},
  {\em J. Math. Phys.} {\bf 23} (1982) 373.

\bibitem{Chandia1997}
O.~Chandia and J.~Zanelli, {\it {Topological invariants, instantons and chiral
  anomaly on spaces with torsion}},  {\em Phys. Rev.} {\bf D55} (1997)
  7580--7585, [\href{http://xxx.lanl.gov/abs/hep-th/9702025}{{\tt
  hep-th/9702025}}].

\bibitem{Zanelli2005}
J.~Zanelli, {\it {Lecture notes on Chern-Simons (super-)gravities. Second
  edition (February 2008)}},
  \href{http://xxx.lanl.gov/abs/hep-th/0502193}{{\tt hep-th/0502193}}.

\bibitem{Garcia2003}
A.~A. Garcia, F.~W. Hehl, C.~Heinicke, and A.~Macias, {\it {Exact vacuum
  solution of a (1+2)-dimensional Poincare gauge theory: BTZ solution with
  torsion}},  {\em Phys. Rev.} {\bf D67} (2003) 124016,
  [\href{http://xxx.lanl.gov/abs/gr-qc/0302097}{{\tt gr-qc/0302097}}].

\bibitem{Achucarro1986}
A.~Achucarro and P.~K. Townsend, {\it {A Chern-Simons Action for
  Three-Dimensional anti-De Sitter Supergravity Theories}},  {\em Phys. Lett.}
  {\bf B180} (1986) 89.

\bibitem{Witten1988a}
E.~Witten, {\it {(2+1)-Dimensional Gravity as an Exactly Soluble System}},
  {\em Nucl. Phys.} {\bf B311} (1988) 46.

\bibitem{Nitti2004}
F.~Nitti and M.~Porrati, {\it {Hidden sl(2,R) symmetry in 2D CFTs and the wave
  function of 3D quantum gravity}},  {\em JHEP} {\bf 01} (2004) 028,
  [\href{http://xxx.lanl.gov/abs/hep-th/0311069}{{\tt hep-th/0311069}}].

\bibitem{Witten2007}
E.~Witten, {\it {Three-Dimensional Gravity Revisited}},
  \href{http://xxx.lanl.gov/abs/0706.3359}{{\tt arXiv:0706.3359}}.

\bibitem{Zamolodchikov1986c}
A.~B. Zamolodchikov, {\it {Irreversibility of the Flux of the Renormalization
  Group in a 2D Field Theory}},  {\em JETP Lett.} {\bf 43} (1986) 730--732.

\bibitem{Lu2010h}
H.~Lu and Z.-L. Wang, {\it {On M-Theory Embedding of Topologically Massive
  Gravity}},  {\em Int. J. Mod. Phys.} {\bf D19} (2010) 1197--1203,
  [\href{http://xxx.lanl.gov/abs/1001.2349}{{\tt arXiv:1001.2349}}].

\bibitem{Blagojevic2006}
M.~Blagojevic and B.~Cvetkovic, {\it {Asymptotic charges in 3d gravity with
  torsion}},  {\em J. Phys. Conf. Ser.} {\bf 33} (2006) 248--253,
  [\href{http://xxx.lanl.gov/abs/gr-qc/0511162}{{\tt gr-qc/0511162}}].

\bibitem{Blagojevic2006b}
M.~Blagojevic and B.~Cvetkovic, {\it {Black hole entropy from the boundary
  conformal structure in 3D gravity with torsion}},  {\em JHEP} {\bf 10} (2006)
  005, [\href{http://xxx.lanl.gov/abs/gr-qc/0606086}{{\tt gr-qc/0606086}}].

\bibitem{Blagojevic2006a}
M.~Blagojevic and B.~Cvetkovic, {\it {Black hole entropy in 3D gravity with
  torsion}},  {\em Class. Quant. Grav.} {\bf 23} (2006) 4781,
  [\href{http://xxx.lanl.gov/abs/gr-qc/0601006}{{\tt gr-qc/0601006}}].

\bibitem{Strominger1998a}
A.~Strominger, {\it {Black hole entropy from near-horizon microstates}},  {\em
  JHEP} {\bf 02} (1998) 009,
  [\href{http://xxx.lanl.gov/abs/hep-th/9712251}{{\tt hep-th/9712251}}].

\bibitem{Cardy1986}
J.~L. Cardy, {\it {Operator Content of Two-Dimensional Conformally Invariant
  Theories}},  {\em Nucl. Phys.} {\bf B270} (1986) 186--204.

\bibitem{Giacomini2007}
A.~Giacomini, R.~Troncoso, and S.~Willison, {\it {Three-dimensional
  supergravity reloaded}},  {\em Class. Quant. Grav.} {\bf 24} (2007)
  2845--2860, [\href{http://xxx.lanl.gov/abs/hep-th/0610077}{{\tt
  hep-th/0610077}}].

\bibitem{Blagojevic2004}
M.~Blagojevic and B.~Cvetkovic, {\it {Canonical structure of 3D gravity with
  torsion}}, in ``Trends in general relativity and quantum cosmology'', volume
  2 (2006) 103,
  [\href{http://xxx.lanl.gov/abs/gr-qc/0412134}{{\tt gr-qc/0412134}}].

\bibitem{Banerjee2010a}
R.~Banerjee, S.~Gangopadhyay, P.~Mukherjee, and D.~Roy, {\it {Symmetries of the
  general topologically massive gravity in the hamiltonian and lagrangian
  formalisms}},  {\em JHEP} {\bf 02} (2010) 075,
  [\href{http://xxx.lanl.gov/abs/0912.1472}{{\tt arXiv:0912.1472}}].

\end{thebibliography}
\providecommand{\href}[2]{#2}\begingroup\raggedright\endgroup

\end{document}